\documentclass[aps,prb,twocolumn,superscriptaddress]{revtex4-2}

\usepackage{times}
\usepackage{changes}

\usepackage{graphicx}
\usepackage{amssymb}
\usepackage{epstopdf}
\usepackage{amsmath}
\usepackage{color}
\usepackage{hyperref}
\usepackage{bbold}
\usepackage{enumitem}
\usepackage[acronym]{glossaries}
\usepackage{physics}
\usepackage[bottom]{footmisc}
\usepackage[version=4]{mhchem}
\usepackage{bm}
\usepackage{ulem}
\usepackage{svg}
\usepackage{mathtools}

\usepackage{txfonts}

\newcommand{\br}{\bm{r}}
\newcommand{\bx}{\bm{x}}
\newcommand{\bn}{\bm{n}}
\newcommand{\bu}{\bm{u}}
\newcommand{\by}{\bm{y}}

\makeatletter

\@ifundefined{textcolor}{}
{
	\definecolor{BLACK}{gray}{0}
	\definecolor{WHITE}{gray}{1}
	\definecolor{RED}{rgb}{1,0,0}
	\definecolor{GREEN}{rgb}{0,1,0}
	\definecolor{BLUE}{rgb}{0,0,1}
	\definecolor{CYAN}{cmyk}{1,0,0,0}
	\definecolor{MAGENTA}{cmyk}{0,1,0,0}
	\definecolor{YELLOW}{cmyk}{0,0,1,0}
}

\graphicspath{{Images/}}

\hypersetup{
	pdfpagemode={UseOutlines},
	bookmarksopen,
	pdfstartview={FitH},
	colorlinks,
	linkcolor={black},
	citecolor={blue},
	urlcolor={blue}}

\usepackage{url}
\usepackage{hyperref}
\hypersetup{
	colorlinks=true,
	linkcolor=black,          % color of internal links
	citecolor=blue,           % color of links to bibliography
	filecolor=magenta,        % color of file links
	urlcolor=cyan
}

\def\QED3{$\text{QED}_3$}

\renewcommand{\Re}{\operatorname{Re}} % For real part
\renewcommand{\Im}{\operatorname{Im}} % For imag part

\newcommand{\iu}{\mathrm{i}} % upright i -- for imaginary numbers
\newcommand{\eu}{\mathrm{e}} % upright e -- for Euler number
\newcommand{\du}{\mathrm{d}} % upright d -- use for differentials
\newcommand{\hc}{\mathrm{h.c.}} % hermitian conjugate

\newcommand{\Ztwo}{\mathbb{Z}_2}
\newcommand{\Z}{\mathbb{Z}}
\newcommand{\Uone}{\mathit{U(1)}}
\newcommand{\SUtwo}{\mathit{SU(2)}}

\newcommand{\SOfive}{\mathit{SO(5)}}

\newcommand{\SO}{\mathit{SO}}
\newcommand{\SU}{\mathit{SU}}

\makeatother	

\begin{document}
	
	\title{Spin-Peierls instability of deconfined quantum critical points}
	
	\author{David Hofmeier}
\thanks{These two authors contributed equally to this work.}
\affiliation{Technical University of Munich, TUM School of Natural Sciences, Physics Department, 85748 Garching, Germany}

    \author{Josef Willsher}
\thanks{These two authors contributed equally to this work.}
	% \email{joe.willsher@tum.de}
\affiliation{Technical University of Munich, TUM School of Natural Sciences, Physics Department, 85748 Garching, Germany}
\affiliation{Munich Center for Quantum Science and Technology (MCQST), Schellingstr. 4, 80799 M{\"u}nchen, Germany}

\author{Urban F. P. Seifert}
\affiliation{Kavli Institute for Theoretical Physics, University of California, Santa Barbara, CA 93109}
\affiliation{Institute for Theoretical Physics, University of Cologne, 50937 Cologne, Germany}

    \author{Johannes Knolle}
\affiliation{Technical University of Munich, TUM School of Natural Sciences, Physics Department, 85748 Garching, Germany}
\affiliation{Munich Center for Quantum Science and Technology (MCQST), Schellingstr. 4, 80799 M{\"u}nchen, Germany}
    \affiliation{Blackett Laboratory, Imperial College London, London SW7 2AZ, United Kingdom}
 
\begin{abstract}
Deconfined quantum critical points (DQCPs) are putative phase transitions beyond the Landau paradigm with emergent fractionalized degrees of freedom.
The original example of a DQCP is the spin-1/2 quantum antiferromagnet on the square lattice which features a second order transition between valence bond solid (VBS) and Néel order. The VBS order breaks a lattice symmetry, and the corresponding VBS order parameter may couple to lattice distortion modes (phonons) at appropriate momenta. We investigate a field-theoretic description of the DQCP in the presence of such a spin-lattice coupling. We show that treating phonons as classical lattice distortions leads to a relevant monopole-phonon interaction inducing an instability towards a distorted lattice by an analogous mechanism to the spin-Peierls instability in one dimension.
Consequently, there is a breakdown of the DQCP which generally becomes a strong first-order transition. 
Taking into account the full quantum nature of the phonons, we argue that the continuous DQCP persists above a critical phonon frequency.
Lastly, we comment on the connection to general gapless, deconfined gauge theories.
\end{abstract}

    \maketitle

\section{Introduction}

Deconfined quantum criticality (DQC) was originally proposed \cite{Senthil2004,Senthil2004B,Senthil2023} as a continuous transition on the square lattice between two ordered states: an antiferromagnetic Néel state and a valence bond solid (VBS).
Such a non-fine-tuned continuous transition between two phases that break independent distinct symmetries is forbidden in the Landau paradigm. Instead, the scenario of DQC relies on the fractionalization of constituent degrees of freedom into spinon excitations interacting with emergent gauge fields which are deconfined \emph{at the critical point}.
While initially studied in the context of quantum magnets, several other systems have been proposed to possibly exhibit DQC \cite{christos20,khalaf21,Liu2022,wang21,liu19,christos2024}.
More generally, deconfined quantum critical points (DQCPs) are understood as a powerful tool to organize competing phases and their transitions \cite{nikolaenko2023,zhang2020}.
While significant progress has been made in identifying low-energy field theories that may describe deconfined quantum critical points and their dualities \cite{PhysRevX.7.031051}, evidence for the realization in a microscopic system is still sought for:
(i) On the numerical side, several salient features of DQC have been observed in simulations of microscopic models or appropriate lattice-regularized field theories, but recent works cast doubt on a scenario of a truly continuous transition and instead point towards weakly first-order behavior, the origin of which currently being under debate \cite{Nahum2015,Poland2019,Ma2020,Nahum2020,DEmidio2023,takahashi2024so5}.
For the purpose of the work at hand, we will nevertheless focus on a scenario of a continuous DQCP, expecting that our results continue to hold upon considering \emph{weakly} first order transitions.
(ii) On the experimental side, systems exhibiting (possibly) continuous Néel--VBS transitions are scarce, with the exception of the candidate material SrCu$_2$(BO$_3$)$_2$ \cite{Cui2023, Guo2023}.
Although this material shows signs of DQCP in a magnetic field, the behavior has recently been shown to turn strongly first order when tuning the lattice under hydrostatic pressure. 

When connecting idealized microscopic models with possible material realizations, experimental realities must be taken into account. Disorder, defects, doping \cite{Kaul2007} and spurious interlayer couplings (in quasi-2d systems) \cite{Lee2019} can have significant impact on the stability and phenomenology of possibly exotic states.
We take this as a motivation to study the stability of the DQCP to spin-lattice coupling.

\begin{figure}
    \centering
    \includegraphics[width=0.9\columnwidth]{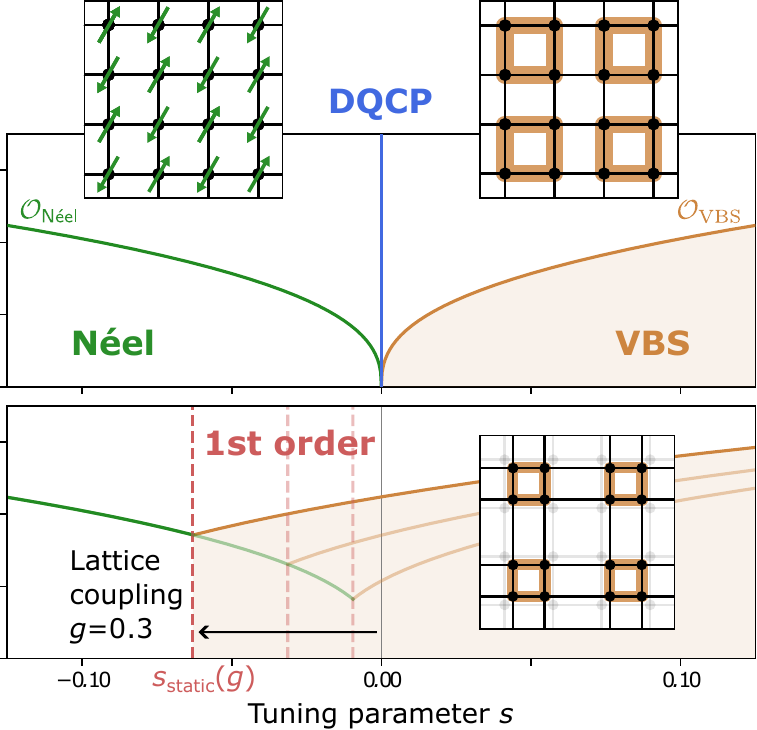}
    \caption{Upper panel: proposed second-order DQCP transition between competing Néel and VBS orders on the square lattice (without lattice coupling $g=0$). Lower panel: spin phonon coupling shifts the transition and renders it first order. The deconfined critical point becomes confined and gapped. Shown here is spin-phonon coupling $g=0.3$ (and lattice stiffness $\mathcal{K}=1$), with $g=0.1,0.2$ in the background.
    }
    \label{fig:fig1}
\end{figure}

The key result of the work at hand is that the coupling to \emph{static} lattice distortion modes renders the deconfined phase transition to exhibit strongly first order behavior, as illustrated in Fig.~\ref{fig:fig1}.
This is in analogy to the Peierls instability, previously discussed mostly for one-dimensional systems of (possibly interacting) fermions/spins coupling to static lattice distortions $u$.
Considering an effective energy functional for such a system, such an instability occurs for infinitesimally small spin-lattice couplings $g > 0$ if the singular response $\mathcal{E}_g\sim (g u)^\chi$ with $\chi < 2$ of the electronic/spin sector outcompetes the (usually) harmonic energy cost (proportional to some lattice stiffness $\mathcal{K}$) of undergoing a distortion, i.e. by minimizing
\begin{equation}
    \mathcal{E}[u] = \mathcal{K} u^2 - (g u)^\chi.
    \label{eq:energycompetition}
\end{equation}
The singular response $\mathcal{E}_g$ is strongly dimension-dependent; a half-filled band of free fermions undergoes the Peierls instability in one dimension but free Dirac fermions are stable in higher dimensions. Less is known about strongly interacting systems in higher dimensions. Recently, some of us have shown that the 2d $U(1)$ Dirac spin liquid (DSL) features such a lattice instability \cite{seifert2023spinpeierls}.
Here, we go beyond quantum critical \emph{phases} to critical points and show that these can also be unstable under a mechanism that resembles the 1d spin-Peierls instability of the Luttinger liquid. This extends works which study the effect of elastic couplings on general quantum critical points \cite{zacharias15,sarkar2023}.
We further go beyond (classical) lattice distortions and consider quantum mechanical phonons with finite frequency $\omega_0$. In this case DQCPs may persist and we derive scaling laws for a parameter regime of stable deconfined criticality 

There are two characteristic features of DQCPs that render the study of couplings to lattice distortion modes a particularly worthwhile endeavor. Firstly, \emph{emergent symmetries} are believed to be common at DQCPs. In the case of the Néel--VBS transition (in 2+1 dim.), this behavior is captured by combining the antiferromagnetic Néel $(n_x, n_y,n_z)$ and VBS $v^a = \eta^i \langle \vec{S}\cdot \vec{S}_a\rangle$ order parameters into the vector $\tilde{n} = (\bm{n}, \bm{v})$, the dynamics of which may be described by a non-linear sigma model (NLSM) supplemented by a topological Wess--Zumino--Witten (WZW) term, which `intertwines' the order parameters. At the critical point, the microscopic $\SU(2) \times C_4$ ($\SU(2)$ spin and fourfold lattice rotation) symmetry is enhanced to an emergent $\SO(5)$ symmetry \cite{nahum15b}. It is thus important to clarify the resilience of such emergent symmetry upon including lattice distortion modes which primarily couple to fluctuating VBS order parameters.

Secondly, the DQCP has an equivalent description in terms of a \emph{deconfined $U(1)$ gauge theory}. As argued by Senthil et al. \cite{Senthil2004}, the Néel order parameter field $\bn$ can be written as a composite object of \emph{fractionalized spinons} which interact with an emergent $\Uone$ gauge field, captured in terms of a CP$^1$ field theory. Formally, the gauge field is compact, rendering instanton events possible, created by monopoles which insert a $2 \pi$ flux of the gauge field.
In this language, VBS order is obtained after the proliferation of said monopoles, which may thus be identified with fluctuating VBS order parameters. The key argument of Ref.~\cite{Senthil2004}, rendering deconfined quantum criticality possible, consists in the fact that such monopoles are forbidden to appear in the low-energy (IR) theory by symmetry. The lowest order term allowed by lattice symmetries corresponds to a quadrupled monopole which however is irrelevant \emph{at the critical point}.
Crucially, these conclusions are based on the symmetry quantum numbers of the monopole operator and are thus lattice dependent \cite{PhysRevLett.111.137202}, the key assumption being that the underlying microscopic Hamiltonian of the system respects the full symmetry of the square lattice.
Lattice distortions, which might be generated by the system spontaneously (and are generally nonzero on the VBS side of the transition) explicitly break these symmetries, calling the underlying assumptions for deconfined criticality in question.

We approach these questions in a low-energy field-theoretical framework, which allows for general statements independent of putative microscopic realizations of DQCP.
As an example of the generality, we present throughout this work a study of the proposed one-dimensional example of DQCP in antiferromagnetic spin chains with anisotropy \cite{Jiang2019,Roberts2019,Mudry2019,Lee2022}.
This makes especially clear the connection with the well-understood Peierls instability of the Luttinger liquid phase.
In 2+1 dimensions, we primarily focus on the canonical example of the Néel--VBS deconfined phase transition, where we show that lattice distortion modes can couple to strongly relevant monopole operators and thereby strongly impact the nature of the phase transition.

The article is structured as follows. In Section \ref{sec:DQCP}, we recapitulate theoretical descriptions of DQC. Then, in Section \ref{sec:DQCP1d} we show how a spin-lattice coupling naturally emerges in a one-dimensional reincarnation of the DQCP. In Section \ref{sec:DQCP2d}, we then turn to the 2d DQCP and argue for its instability upon coupling to classical lattice distortion modes. In Section \ref{sec:Dynamics} we show how quantizing the lattice excitations alleviates this instability for weak couplings, and derive the critical scaling relation between the spin-lattice coupling and the phonon frequency below which the lattice remains stable. Finally, we conclude by commenting on the connections to novel models of DQCP, deconfined phases with fermions, and the recently proposed Stiefel liquids \cite{Zuo2021}.

\section{Deconfined Quantum Criticality}
\label{sec:DQCP}
An effective low-energy theory for quantum antiferromagnets (in d+1 dimensions)
is the $O(3)$ NLSM for antiferromagnetic order parameter fluctuations \cite{Haldane1988},
\begin{equation} \label{eq:O3-NLSM}
    S_{O(3)} = \frac{1}{2e^2}\int_0^\beta \dd{\tau} \int \dd[d]{\bx} \left[ v^2(\nabla\bn)^2 + (\partial_\tau\bn)^2\right] + S_{\text{top}}[\bn]
\end{equation}
where $\bn(\br)$ is the continuum field corresponding to the Néel order parameter $\bn_j = \eta_j \langle\bm{S}_j\rangle$, where $\eta_j=\pm1$ alternates on neighboring sites $j$ of the lattice
(we set the spin-wave velocity $v=1$ in the following).
Here, $S_{\text{top}}[\bn]$ is a topological term that follows from the Berry phase contribution of the spin path integral \cite{Sachdev2011,Fradkin2013}.
Explicitly working on a lattice, it is given by
\begin{equation}
    S_{\text{top}}[\bn] = \frac{i}{2}\sum_{j} \eta_{j}\int \dd{\tau} \int_0^1 \dd{u} \left[\bn_j\cdot(\partial_u\bn_j\times\partial_\tau\bn_{j})\right]
    \label{eq:O3top}
\end{equation}
with $\bn(\tau)$ smoothly extended to $\bn(\tau,u)$ such that $\bn(\tau,u=0) = (0,0,1)^T$ and $\bn(\tau,u=1) = \bn(\tau)$. As pointed out in \cite{Read1989,Senthil2004}, the topological term is of vital importance; because of it, the NLSM does not flow to a disordered phase at strong coupling but rather, both in 1d and 2d, a VBS phase is realized due to the proliferation of topological defects in $\bn$. 

While the possibility of a direct continuous phase transition between Néel and VBS order follows from the $O(3)$ NLSM, the nature of the transition is difficult to access in this framework, as it is fundamentally an expansion about the Néel ordered phase.
Instead, one may turn to field theories where intertwined Néel and VBS order parameters are treated on a more equal footing. This includes a ``superspin'' NLSM with additional Wess--Zumino--Witten terms \cite{Senthil2006} and the non-compact $CP^1$ gauge theory that describes deconfined spinons at the critical point \cite{Senthil2004}, which we review below. Note that dualities between several such theories are thought to exist \cite{PhysRevX.7.031051}.

\subsection{Deconfined Quantum Criticality in One Dimension}

An insightful analog to two-dimensional deconfined quantum critical phenomena are one-dimensional models, which have recently received renewed attention \cite{Jiang2019,Roberts2019,Mudry2019,Yang2023,Lee2022,Romen2024}. Here, we will focus on the scenario of a transition between Néel order along the $\hat{z}$-axis and a VBS-ordered phase, where a finite anisotropy easy-axis anisotropy (reducing the spin rotation symmetry $\SU(2) \to U(1) \times \mathbb{Z}_2$) is required to stabilize long-range magnetic order.
To describe the transition itself, one may unify the Néel vector $\bn$ and VBS order parameter $v_j = \eta_j\langle\bm{S}_j\cdot\bm{S}_{j+1}\rangle$ into a four-component vector $\Tilde{\bn} = (\bn,v)$ whose dynamics is governed by an $\mathrm{O}(4)$ NLSM supplemented by a Wess--Zumino--Witten (WZW) term \cite{Senthil2006, Nahum2020} as well as appropriate anisotropies,
\begin{equation}
    S_1^{\mathrm{DQCP}}[\tilde{\bm{n}}] = S_{1}^{\mathrm{kin}}[\tilde{\bm{n}}] + \int \dd[2]{x}s(\bm{n}^2-v^2) + \Gamma_1^{SO(4)} + S_\mathrm{aniso}.
    \label{eq:GenNLSM1d}
\end{equation}
Here, $s$ is a tuning parameter for the transition between Néel ($s<0$) and VBS ($s>0$), and we explicitly add the anisotropy $S_\mathrm{aniso} = \alpha_{\mathrm{aniso}} \int \du^2 x (n_3^2 + v^2)$ to stabilize the relevant ordered phases to realise a DQCP. Finally, 
\begin{align}
   S_{d}^{\mathrm{kin}}[\tilde{\bm{n}}] &=  \frac{1}{2e^2}\int \dd[d+1]{x}(\nabla \Tilde{\bn})^2 \quad \text{and}\\
    \Gamma_1^{\SO(d+3)} &= \frac{2\pi i}{\Omega_{d+3}}\epsilon_{j_1\dots j_{d+3}}\int \dd[d+1]{x}\int_0^1 \dd{u}\, \Tilde{n}_{j_1}\partial_{x_1}\Tilde{n}_{j_2}\dots\partial_u\Tilde{n}_{j_{d+3}}
\end{align}
are kinetic and topological WZW terms in $d$ spatial dimensions. This final term is crucial for the physics of the transition; in the AFM phase it recovers the $O(3)$ NLSM's topological term \eqref{eq:O3top}, which in 1d can be rewritten as
\begin{equation}
    S_{\mathrm{top}} = \frac{i}{4}\int \dd{x}\dd{\tau} \bn\cdot(\partial_x\bn\times\partial_\tau\bn)  = i\pi Q_{x\tau}
    \label{eq:Stop1d}
\end{equation}
in terms of the integer Pontryagin index $Q_{x,\tau} \in \mathbb{Z}$.
The topological angle $\pi$ guarantees that the transition at $s=0$ is gapless \cite{PhysRevB.36.5291,PhysRevB.43.3337}.
Additionally, the WZW term is responsible for topological defects in the VBS phase carrying a $S=1/2$ degree of freedom, making them seeds for Néel order and conversely causes the topological defects in the Néel-ordered phase to transform non-trivially under lattice symmetries such that their condensation leads to VBS order.
This so-called \emph{intertwinement} of order parameters is a key underlying feature of the DQCP \cite{Levin2004,Tanaka2005,Senthil2006,Metlitski2018}. 

Assuming a sufficiently strong easy-axis anisotropy the NLSM supervector $\Tilde{\bn}$ can be parameterized by a single angle $2\phi$ and the effective action becomes
\begin{equation}
    S_{\mathrm{SG}} = \int \dd{x}\dd{\tau} \left[\frac{1}{2\pi K}\left((\partial_x\phi)^2+(\partial_\tau\phi)^2\right)+\mu\cos(4\phi)\right] \,,
    \label{eq:SG}
\end{equation}
called the Sine-Gordon model where $K$ is the Luttinger parameter. Such a theory can also be derived from appropriate microscopic models via Abelian bosonization \cite{Giamarchi2003,Sachdev2011}. For $\mu>0$, the model flows to a VBS ordered-state, while for $\mu<0$, it has Néel order.
As discussed recently, the critical line at $\mu=0$ for $1/8 < K < 1/2$ can be understood as a line of deconfined quantum critical points with emergent $U(1)\times U(1)$ symmetry \cite{Mudry2019,Roberts2019,Jiang2019}. This will be our regime of interest.
The respective order parameters are given by $n_3 \sim \cos 2 \phi$ and $v \sim \sin 2 \phi$, both having scaling dimension $K$.
The order parameters onset with power-law behavior $n_3,v \sim |\mu|^{\beta}$ governed by the exponent $\beta = \nu_\mu K$, where $\nu_\mu^{-1} = 2-4K$ in 1d.
The transformation properties of $\phi$ and the order parameters under internal spin rotations and the external lattice symmetries can be found in Table \ref{tab:SymSG}.

\subsection{Deconfined Quantum Criticality in Two Dimensions}

We now return to the 2+1-dim. case of putative DQCP on the square lattice. Here, the analogue of the NLSM with a WZW term takes the form \cite{Tanaka2005,Senthil2006}
\begin{equation}
    S_2^{\mathrm{DQCP}}[\tilde{\bm{n}}] = S_{2}^{\mathrm{kin}} [\tilde{\bm{n}}] + \int \dd[3]{x} s(\bm{n}^2-\bm{v}^2) + \Gamma_2^{\SO(5)}[\tilde{\bm{n}}],
    \label{eq:GenNLSM}
\end{equation}
where the three-component Néel order parameter $\bn$ and the two-component VBS order parameter $v_j^a = \eta_j\langle\bm{S}_j\cdot\bm{S}_{j + \hat{\br}_a}\rangle$ ($a=x,y$) have been combined into 5-component vector $\tilde{\bn}$.
The 2d WZW term $\Gamma_2^{\SO(5)}$ reduces to the spin-1/2 Berry phase in the case of a single $\mathbb{Z}_4$ vortex within the VBS phase \cite{Levin2004}, again being the source of intertwinement in this theory \cite{Levin2004,Tanaka2005,Senthil2006}.

In the case of the two-dimensional DQCP, seminal works \cite{senthil04b,Senthil2004} have argued that the transition may be described in terms of fractionalized spinons $z$ interacting with a $\Uone$ gauge field $a_\mu$ within a $CP^1$ model of the form
\begin{equation}
    \mathcal{L}_c = \sum_{\alpha=1}^2\abs{(\partial_\mu-ia_\mu)z_\alpha}^2 + r\abs{z}^2 + u(\abs{z}^2)^2+\lambda_4\left[\Phi^4+(\Phi^\dagger)^4\right].
    \label{eq:NCCP1}
\end{equation}
The relevant boson-mass parameter $r$ tunes the transition and we have additionally included possible anisotropies $u$ which preserve $\SU(2)$ spin symmetry.
Here, the Néel order parameter field is to be identified with a composite object of spinons via the Hopf map $\bn = z^\dagger \bm{\sigma}z$, and $\Phi^\dagger$ is a $\Uone$ monopole insertion operator. 
From this mapping, it can be shown that the total flux of the $\Uone$ gauge field is related to the instanton number of the Néel order parameter field \cite{Auerbach1994}, and hence $\Uone$ monopole insertions in the CP$^1$ model correspond to two-dimensional topological hedgehog defects in the NLSM.

These monopoles transform with non-trivial phases under the lattice symmetries \cite{Read1990,Lee2019}.
The existence of the topological term \eqref{eq:O3top} means that hedgehogs of $\bn$ located at different sites contribute with a different phase factor. 
In the $CP^1$ formulation this has to be accounted for and the monopoles $\Phi^\dagger$ will pick up non-trivial phases under lattice symmetries. The precise symmetry properties are summarized in Table \ref{tab:SymMonopole}, and will be of importance when searching for appropriate monopole-lattice couplings. Using the behavior of $\Phi^\dagger$ under translation symmetry, we can also identify it with the VBS order parameter
\begin{equation}
    v_1 \sim \Re{\Phi} = \Phi_1 \quad \text{and}\quad v_2\sim \Im{\Phi} = \Phi_2\,.
\end{equation}
Therefore, the lowest order symmetry-allowed monopole term is $(\Phi^\dagger)^4$.
It has been established in exact diagonalization that this $8\pi$-monopole is irrelevant at the transition \cite{zhou2024,takahashi2024so5}, meaning the $\Z_4$ lattice rotation symmetry is enhanced to $U(1)$. This $\SUtwo\times \Uone$ symmetry may be further enhanced to $\SOfive$ by tuning $s$ (or equivalently $r$ in the NCCP1 model).
The DQCP correlation length exponent $\nu_s = 1/(3-\Delta_s) = 1.38$ and $2\pi$-monopole scaling dimension $\Delta_\Phi\sim0.61$ have also been measured numerically \cite{takahashi2024so5}, leading to an order parameter onset in 2d DQCP given by $\beta = \nu_s \Delta_\Phi$ (as plotted in the upper panel of Fig.~\ref{fig:fig1}).

\section{Spin-Lattice Coupling in One Dimension}
\label{sec:DQCP1d}
\subsection{Spin-Lattice Coupling}

Before we turn to the most prominent example of DQC, we focus on the 1d case \cite{Jiang2019,Roberts2019,Mudry2019,Lee2022}. Here, we make use of low-energy field theory arguments, largely following previous discussions in the context of one-dimensional Heisenberg chains (without anisotropy) \cite{Citro2005}.
To this end, we note that a static lattice distortion $u_n$ can couple to spin bilinears via a Hamiltonian of the form $H_{\mathrm{sp}} = \frac{g}{2} \sum_n u_n\bm{S}_n\cdot\bm{S}_{n+1}$. As previously discussed in the context of one-dimensional Heisenberg chains (without anisotropy) \cite{Citro2005}, at lowest energies only staggered lattice distortions (i.e.~with lattice momentum $Q=\pi$) are important.
Writing $u_n = (-1)^n \delta u$ and bosonizing, the distortion field can be seen to couple to the field $\phi$ in the sine-Gordon action \eqref{eq:SG} via
\begin{equation}
    H_{\mathrm{sp}} \sim h\, \sin(2\phi)\,,
\end{equation}
where $h = g\, \delta u$.
We emphasize a complementary symmetry-based top-down construction of the spin-lattice coupling: Equipped with the symmetry transformation behaviors of the field $\phi$ in the Sine-Gordon model in Table~\ref{tab:SymSG} we deduce that the $\sin 2\phi$ interaction, which is odd under translations $T_x$ and inversion $I$, has a symmetry-allowed coupling to a staggered lattice distortion field $h = g\, \delta u$.
\begin{table}
    \centering
        \(\begin{array}{c||c | c| c || c}
        \text{Symmetry} & \phi& n_3 & v & \delta u\\ \hline
        U_1^{xy} & \phi & n_3 & v & \delta u\\
        \mathbb{Z}_2^z & -\phi + \pi/2& -n_3 & v & \delta u\\
        T_x & \phi+\pi/2& -n_3 & -v& -\delta u\\
        \mathcal{I} & -\phi & n_3 & -v & -\delta u \\ 
        \mathcal{T} & -\phi + \pi/2 & -n_3 & v & \delta u
        \end{array}\)
    \caption{The symmetry properties of the field $\phi$ under the internal spin symmetries $U(1)^{xy}$ and $\mathbb{Z}_2^z$, as well as the external lattice symmetries: translation $T_x$ and bond-inversion $\mathcal{I}$ as is similarly obtained in bosonization works \cite{Giamarchi2003,Lee2019}. The order parameters are defined $n_3 \sim \cos(2\phi)$ and $v\sim \sin(2\phi)$. The transformation of the staggered distortion field $u_i = (-1)^i \delta u$ is written for comparison; it is a singlet under internal spin-rotation symmetries.}
    \label{tab:SymSG}
 \end{table}
Thus the field theory for the DQC upon coupling to lattice distortion modes becomes
\begin{equation}
    S_u = S_{\mathrm{SG}} + \int \dd{x}\dd{\tau} \lambda \sin(2\phi)\,,
\end{equation}
where we define the field theory coupling $\lambda\sim h$.
The perturbative RG equations of this model read \cite{PhysRevB.22.1405}
\begin{subequations}
    \begin{align}
        \beta_K 
        % = \dv{K}{\ell} 
        &=  -K^2 (\lambda^2+4\mu^2)\,,\label{eq:betaK}\\
        \beta_{\lambda} 
        % = \dv{\lambda}{\ell} 
        &= \left(2-K\right)\lambda + \frac{2}{u}\lambda\mu \, K\,,\label{eq:beta2}\\ 
        \beta_{\mu} 
        % = \dv{\mu}{\ell} 
        &= (2-4K)\mu + \frac{1}{2u}\lambda^2K \label{eq:beta4} \,. 
    \end{align}
\end{subequations}
No real-valued fixed points for the full three dimensional RG flow are found.
We note that this is in contrast to previous works which predict a second-order Ising transition at finite $\lambda$ \cite{PhysRevB.22.1405,Delfino1998}.
This apparent disparity stems from the fact that the analyses in Refs.~\onlinecite{PhysRevB.22.1405,Delfino1998} did not account for the flow given by Eq.~\eqref{eq:betaK} of the Luttinger parameter $K$:
assuming $K$ to be constant, we indeed find two new fixed points for \eqref{eq:beta2} and \eqref{eq:beta4}. However, these fixed points do not persist in the full three-dimensional RG flow.
From Eqs.~\eqref{eq:betaK}--\eqref{eq:beta4} we conclude that the RG flow for any finite $\lambda \neq 0$ at the original DQCP line $(1/8<K<1/2,\mu=0)$ is non-zero, implying the absence of deconfined quantum criticality in the presence of a finite lattice distortion.
\subsection{Lattice Instability}

In the absence of spin-lattice couplings, the undistorted lattice (i.e. $h \equiv 0$) minimizes the system's elastic energy
\begin{equation}
    E_{\text{cost}} = N(\mathcal{K}/g^2)h^2\,,
\end{equation}
where $\mathcal{K}$ is the inherent stiffness of the lattice and $N$ the number of sites.
A spontaneous distortion and thus instability of the DQCP will only occur if the energy gain due to a gap-opening in the spin sector out-competes the energy cost associated with distorting the lattice. This energy gain scales with the gap as
\begin{equation}
    E_{\text{gain}}\sim N\Delta^2\,.
\end{equation}
and in higher dimensions this will generalize to ${E_{\text{gain}}\sim N\Delta^{d+1}}$.
How does the gap $\Delta$ scale with the distortion? This follows from a straightforward RG argument, see e.g. \cite{Giamarchi2003}.
First, note that for $\abs{\lambda}\geq1$, the perturbative RG breaks down and the size of the gap can simply be determined by considering fluctuations $\delta\phi$ about the classical ground state $\phi_0 = \frac{\pi}{4}$. (Without loss of generality, we consider the $\lambda\to-\infty$ ground state.)
In this regime, it is justified to expand, $\phi = \phi_0 +\delta \phi$, yielding
\begin{equation}
    \sin(2\phi) = 1 - 2(\delta\phi)^2 + \order{(\delta\phi)^4}\,.
\end{equation}
So, for $\abs{\lambda} \geq 1$, the $\sin(2\phi)$-interaction simply generates a mass term, resulting in a constant gap $\Delta_0$. With $\Delta_0$ corresponding to the gap at $\lambda(s_c) \sim 1$, we use $\Delta(s) = e^{-s_c} \Delta_0$ and the RG flow of $\lambda$ \eqref{eq:beta2} to obtain the scaling of the gap as
\begin{equation}
    \Delta = \Delta_0\lambda^{1/(2-K)} 
    \,.
\end{equation}

Within our approximation, the total energy of the system is then  composed of a (harmonic) energy cost of the distortion, and an energy gain due to the opening of the gap, 
\begin{equation} \label{eq:Etot1d}
    \mathcal{E} = \frac{1}{N}\left(E_{\text{cost}}-E_{\text{gain}}\right) 
    = (\mathcal{K}/g^2) h^2 - c h^{2/(2-K)},
\end{equation}
implying an instability towards a lattice distortion for $K<1$. The marginal case $K=1$ recovers the Peierls instability of free fermions, and $K=1/2$ the spin-Peierls instability of the pure Heisenberg chain.

Of interest for DQC is the region $1/8<K<1/2$; here the lattice is always unstable towards this distortion at the original DQCP $\mu=0$. The previous analysis implies an equilibrium VBS order, a lattice distortion $h_0$, and a corresponding gap $(g^2/\mathcal{K})^{1/(2-2K)}$. From this, it follows that (i) deconfined criticality is spoiled, i.e. the Néel--VBS transition becomes a first order transition, and (ii) the location of the critical point is shifted. Tuning $\mu<0$ in the pure DQCP model opens up an Néel gap which scales as $\mu^{1/(2-4K)}$; when the energy gain due to this gap opening is equal to the aforementioned energy gain which drives the spin-Peierls gap at the original DQCP, there will be a first-order transition. This direct VBS to Néel order transition therefore occurs at the shifted value
\begin{equation}
    \mu_{\mathrm{static}} = -(g^2/\mathcal{K})^{(2-4K)/(2-2K)}.
\end{equation}
We exclude the possibility that the shifted transition remains deconfined, since no sort of second order Néel--VBS transition can exist in 1d on a lattice without translational symmetry.
This effect can be elucidated by considering the 1d topological term Eq.~\eqref{eq:Stop1d}
in the presence of a finite distortion $\delta u$, $S_{\text{top}} = i\pi (1+\delta u/J) Q_{x\tau}$, which can be seen by using the Haldane expansion for the Heisenberg spin chain in the presence of an additional \emph{staggered interaction} induced by a lattice distortion, see also \cite{AFFLECK1986409}.

\section{Spin-Lattice Coupling in Two Dimensions}
\label{sec:DQCP2d}
\subsection{Spin-Lattice Coupling}
We observe that, just like in 1d, there exists an operator that is relevant at the DQCP but disallowed in the field theory due to its non-trivial behavior under lattice symmetries. The role of $\sin(2\phi)$ in 1d is taken over by the monopole operator $\Phi^\dagger$. Its transformation behavior has been obtained by explicitly transforming the expression for the Berry phase in the presence of a single monopole \cite{Read1989,Lee2019}.
Table \ref{tab:SymMonopole} summarizes their symmetry properties. 
\begin{table}
\centering
\(\begin{array}{c|| c | c||c}
\text{Symmetry} & \bm{n} & \Phi^\dagger & u^*\\ \hline\hline
\SUtwo^{\mathrm{spin}} &R_{ab} \bm{n}_b & \Phi^\dagger & u^* \\
T_x & -\bn & -\Phi & -u\\
T_y & -\bn & \Phi & u\\
R_{\pi/2}^{\text{site}} & \bn & i\Phi^\dagger & iu^* \\
\sigma_x & \bn & \Phi & u\\
\mathcal{T} & -\bm{n} & \Phi & u
\end{array}\)
\caption{Symmetry properties of the monopole operator $\Phi^\dagger$ \cite{PhysRevB.42.4568,Lee2019} and the phonon mode $u$ under the elementary lattice symmetries of the square lattice. The transformation of the lattice distortion field $u = u_1 + \iu u_2$ is written for comparison. We include the action of spin rotation on the order parameters: the Néel order transforms under the representation $R\in\SO(3)$, and the VBS order is a singlet, i.e., the monopole is left invariant (this latter fact is also true with possible anisotropy).}
\label{tab:SymMonopole}
\end{table}
Again, we can now ask whether there exists a lattice distortion that transforms the same way such that the product of a monopole and a distortion would be symmetry allowed.
To find this, we recall the definition of the VBS order parameters $v^a(\bm{r}_i) = \eta_i \langle \vec{S}_{\bm{r}_i}\cdot \vec{S}_{\bm{r}_i+\hat{\bm{r}}_a}\rangle$ where $\hat{\bm{r}}_{a=1,2} = \hat{\bm{x}},\hat{\bm{y}}$ are the lattice unit vectors.

Defining a general distortion field $\bm{u}(\bm{x})$ which shifts the lattice sites $\br_i\to \bm{x} =  \br_i+\bm{u}(\bm{x})$,
we expand in a Fourier basis $\bm{u}(\bm{x}) = \sum_{\bm{Q}} \bm{u}^a_{\bm{Q}} \eu^{\iu \bm{Q} \cdot \bm{x}}$, then we extract the two longitudinal components 
\begin{equation}
    u_a = \hat{\bm{Q}}_a \cdot \bm{u}_{\bm{Q}_a},\quad
    \bm{Q}_a = (\pi,0),\, (0,\pi).
\end{equation}
Assuming that the distortion field at wavevectors $\bm{Q}_1$, $\bm{Q}_2$ is purely longitudinal, we can thus write in real space
\begin{equation}
    \bu(\bx) = \begin{pmatrix} u_x(\bx) \\ u_y(\bx) \end{pmatrix} = \begin{pmatrix}
        1 \\ 0
    \end{pmatrix} \eu^{\iu \pi x_1} u_1 + \begin{pmatrix}
        0 \\ 1
    \end{pmatrix} \eu^{\iu \pi x_2} u_2.
    \label{eq:DistFieldGeneral}
\end{equation}
We may combine these two contributions into a complex field $u^* = u_1 + \iu u_2$ as is done for the VBS order parameters to form the monopole $\Phi \sim v_1 + \iu v_2$.
We explicitly evaluate the transformation properties of this field $u_a$ under transformations of the lattice $r_a \to g(r_a) = M_{ab}r_b + \delta_a$. The components transform under the symmetry group of the lattice actively as $u_a(\bx) \to M_{ab}u_b(g^{-1}\bx)$; we summarize the results in \ref{tab:SymMonopole}, which shows that this field transforms identically to the VBS monopoles.
We conclude that the monopole-lattice coupling
\begin{equation}
    \mathcal{O}_{mp} \sim u \, \Phi^\dagger + \hc \sim u_1v_1 + u_2v_2
    \label{eq:coupling}
\end{equation}
is therefore symmetry-allowed.
This appears in the action through the coupling $S_{mp} = -g \int\dd[3]{x} \mathcal{O}_{mp}(x)$
On the level of the NLSM, the lattice couples linearly to the VBS order parameters
\begin{equation}
    S_2[\tilde{\bm{n}},\bm{h}] = S_2^{\mathrm{kin}} [\tilde{\bm{n}}] + \int \dd[D]{x} \left[s(\bm{n}^2-\bm{v}^2) - \bm{h}\cdot \Tilde{\bm{n}}\right]
    + \Gamma_2^{\SO(5)}[\tilde{\bm{n}}]\, 
    \label{eq:nlsm_with_h_field}
\end{equation}
with $\bm{h} \sim g (0,0,0,u_1,u_2)^T$. 
This explicitly breaks the emergent $\SO(5)$ symmetry and will have the effect of pinning the VBS components of the topological NLSM field $\tilde{\bm{n}}$ to the lattice distortion field $\bm{h}$. We expect this term to make the transition tuned by $s$ first order whenever $\bm{h}$ takes a non-zero value.

\subsection{Lattice Instability}
We are interested in understanding the gap induced by the perturbation $\bm{h}$ when the NLSM of Eq.~\eqref{eq:nlsm_with_h_field} is tuned to the strong-coupling fixed point.
In analogy with the result in 1d, the gap scaling as a function of distortion is controlled by the conformal data at this fixed point
\begin{equation}
    \Delta(\bm{h}) \sim |\bm{h}|^{\nu_\Phi}
\end{equation}
with $\nu_\Phi = 1/(3-\Delta_\Phi) = 0.419$ the correlation length exponent of the monopole operator \cite{takahashi2024so5}.
We recall that the distortion $\bm{h}$ in \eqref{eq:nlsm_with_h_field} is a (vector-valued) classical parameter that represents a static, non-local distortion of the lattice (we treat fluctuations of this field in the next section).
Formally, the free energy of the coupled system may be written as $ \mathcal{F}(\bm{h}) = (\mathcal{K}/g^2) |\bm{h}| - \log \int\!\mathcal{D}\tilde{\bm{n}}\, \eu^{-S_2[\tilde{\bm{n}},\bm{h}]}$. The saddle-point behavior of the lattice distortion field is determined by the respective scaling of the energy cost of a distortion and the induced energy gain.
We tune $s$ to the gapless DQCP with an emergent $\SO(5)$ symmetry and, in a result analogous to 1d [in Eq.~\eqref{eq:Etot1d}], we find the saddle point behavior of the distortion field is governed by the effective potential
\begin{equation}
    \mathcal{E}(\bm{h}) = (\mathcal{K}/g^2)\abs{\bm{h}}^2 - c\abs{\bm{h}}^{3 \nu_\Phi}\,.
    \label{eq:2dEnergyDensity}
\end{equation}
Hence, we conclude that a finite spontaneous equilibrium distortion $\bm{h}_0 \neq 0$ is induced at the DQCP on the square lattice.
The condition for a lattice instability in 2d reads $3\nu_\Phi < 2$ or $\Delta_\Phi < 3/2$, as is believed to be satisfied for the DQCP.

The shifted transition must be rendered first order by this distortion, since the coupling $-\bm{h}\cdot \Tilde{\bm{n}}$ will always destroy the critical point. That is to say that on an explicitly distorted lattice, no DQCP can be present since both the strongly relevant single and double monopole $\Phi^\dagger/(\Phi^\dagger)^2$ become symmetry-allowed \cite{Nahum2011} (this is stronger than the case of non-staggered rectangular-lattice distortions where translational symmetry is preserved \cite{PhysRevLett.111.137202,Wang2017,Metlitski2018}).
As for 1d, we also see that the transition is shifted away from the original DQCP and into the original Néel phase. The energies of the VBS and Néel phases cross when the $s$-induced gap to VBS excitations in the Néel phase becomes equal to the distortion-induced gap $\Delta(\bm{h}_0)$ in the VBS phase. This produces a modified transition point
\begin{equation}
    s_{\mathrm{static}} = - [g^2 / \mathcal{K}]^{(3-\Delta_s)/(3-2\Delta_\Phi)}.
    \label{eq:sstatic}
\end{equation}

\begin{figure}
    \centering
    \includegraphics[width=.9\columnwidth]{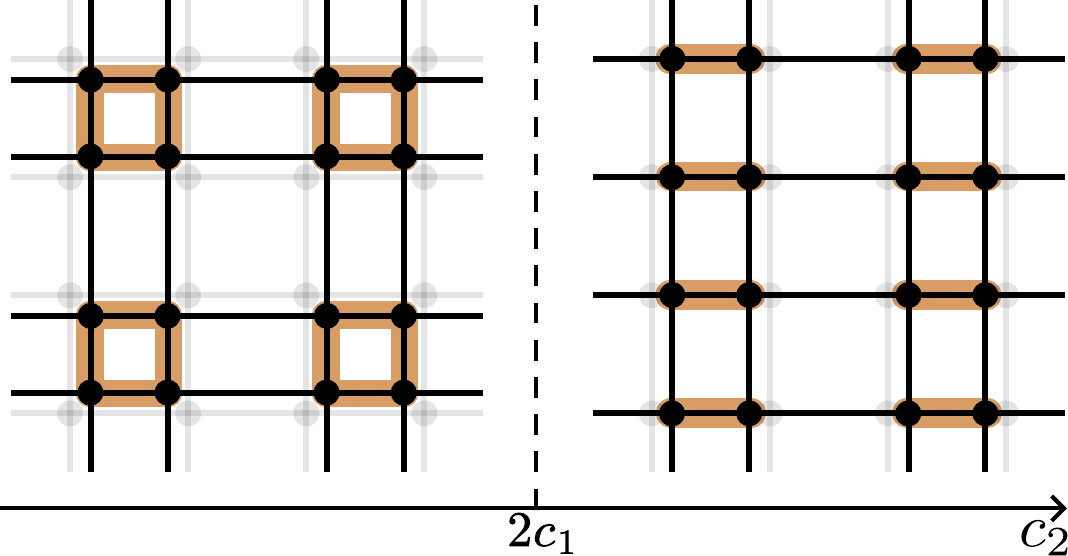}
    \caption{ Possible lattice distortions of the square lattice, depending on the microscopic constant $c_2$. The ground state of the system has the corresponding VBS order, i.e. pVBS for $c_2<2c_1$ and cVBS for $c_2>2c_1$.}
    \label{fig:LatticeDist}
\end{figure}

From the energy density \eqref{eq:2dEnergyDensity}, it would naively follow that there are infinitely many degenerate distorted ground states. It should be noted, however, that we have neglected so far higher order symmetry-allowed terms in the lattice potential. Taking these into account, the energy density up to fourth order reads
\begin{equation}
    \mathcal{E}'(\bm{h}) = \mathcal{E}(\bm{h}) + c_1(h_1^4+h_2^4)+c_2(h_1^2h_2^2)\,.
\end{equation}
To ensure stability, we must require $c_1>0$ and $c_2 > -2c_1$. The coefficient $c_1$ reduces the unphysical $U(1)$-symmetry to a more physical $\mathbb{Z}_4$-symmetry and $c_2$ ultimately determines the type of distortion. For $c_2>2c_1$, the energy density is minimized for a distortion in columnar direction (with either $h_{1,2}=0$) and for $c_2<2c_1$ the distortion is towards the plaquette centers (with $h_1=h_2=h$). Accordingly, the system enters a columnar VBS (cVBS) or plaquette VBS (pVBS) phase, respectively. The value of $c_2$ seems to be a microscopic property (determined by both the lattice dynamics and induced by spin interactions at the new strong-coupling fixed point) and we are not aware of any methods for accessing this from our top-down symmetry approach. To accurately specify the exact distortion, one must most likely resort to numerical methods. We summarize the different distortions in Figure \ref{fig:LatticeDist}.

We briefly note that the energetic analysis provided by \eqref{eq:2dEnergyDensity} represents a `strong-coupling' approach to the coupled spin-lattice system. This describes the system in the thermodynamic limit at zero temperature which, in realizations of DQCP, may only exist below some finite length scale. For example, in numerical simulations the finite-system size $L$ is a relevant perturbation, or in experiments which necessarily have a finite inverse temperature $\beta\sim T^{-1}$.
A complementary `weak-coupling' calculation (of the related DSL) \cite{seifert2023spinpeierls} suggests that the instability sets in below a temperature $T_{\mathrm{SP}} \sim g^{3-2\Delta_\Phi}$.

\section{Adding Dynamics}
\label{sec:Dynamics}
Moving beyond the adiabatic approximation of the lattice, we will now develop a theory of dynamical optical phonons coupled to the DQCP. This approach has the advantage that it more accurately captures the physics of the lattice phonons which generally can be expected to have a finite energy $\omega_0>0$ at nonzero (lattice) momenta.

\subsection{Phonons in One Dimensions}
From the static case, we know that the spins will couple to phonons with momentum $\pi$ at the edge of the Brillouin zone.
Hence, we now promote the Fourier component of the displacement field with lattice momentum $\pi$ to be dynamical, which we henceforth for simplicity denote $u(x,\tau)$.
Then, the combined action of spins and phonons reads
\begin{equation}
    S(\omega_0) =  S_{\mathrm{SG}} + \int \dd{x}\dd{\tau}\left\{ \frac{\rho}{2}\left[|\partial_\tau u|^2 + \omega_0^2|u|^2\right] + 2u\sin(2\phi)\right\}
    \label{eq:fullS1d}
\end{equation}
with the phonon mass density $\rho = m/a$ and frequency $\omega_0 \sim \omega(\pi)$. Integrating out the quadratic displacement field gives
\begin{align}
    \begin{split}
    S(\omega_0) &= S_{\mathrm{SG}} + g^2\int \dd{x}\dd{\tau} \dd{\tau'} G_{\omega_0}(\tau-\tau')\\&\times\left[\cos(2(\phi+\phi'))-\cos(2(\phi-\phi'))\right]
    \label{eq:SGafterIntOut}
    \end{split}
\end{align}
with the phonon propagator 
\begin{equation}
    G_{\omega_0}(\tau-\tau') = \frac{1}{2\rho\omega_0} e^{-\omega_0\abs{\tau-\tau'}}\,.
\end{equation}
For notational brevity, we drop the space-time dependence of $\phi$ whenever it is clear and define $\phi' = \phi(\br,\tau')$. A similar action, without the $\cos(4\phi)$ interaction in $S_{\mathrm{SG}}$, has previously been derived in \cite{Citro2005} to describe the spin-Peierls transition of the $\SUtwo$-symmetric Heisenberg model. Here, we analyze its meaning in the context of DQC where the $\cos(4\phi)$ interaction is relevant.
Let us begin by considering the two limits $\omega_0\to0$ and $\omega_0\to\infty$, called the adiabatic and antiadiabatic limit, respectively \cite{Giamarchi2003}. The former is simply the static case from above. There, we concluded that the deconfined quantum critical transition gets replaced by a first order transition. 

Moving on to the antiadiabatic limit,  we can use the representation of the delta-function (with appropriate normalization)
\begin{equation}
    \frac{\omega_0}{2}e^{-\omega_0\abs{\tau-\tau'}} \overset{\omega_0\to\infty}{\longrightarrow} \delta(\tau-\tau')
    \label{eq:AntiadiabaticDelta}
\end{equation}
to find that the dynamical phonons simply generate another $\cos(4\phi)$ interaction and we have
\begin{align}
    \begin{split}
    S(\omega_0\to\infty) = \int \dd{x}\dd{\tau} \bigg\{&\frac{1}{2\pi K} \left[(\partial_x\phi)^2 + (\partial_\tau\phi)^2\right]\\&+\mu_{\text{new}}\cos(4\phi)\bigg\}
    \end{split}
\end{align}
with a new coupling
\begin{equation}
    \mu_{\text{new}} = \mu+\frac{g^2}{\rho\omega_0^2}\,.
\end{equation}
This means that DQC persists and the critical line simply gets shifted from $\mu=0$ to 
\begin{equation}
    \mu_c = -\frac{g^2}{\rho\omega_0^2} \,.
\end{equation}
Clearly, the model shows qualitatively different behavior in the two opposite limits and so we can expect some critical frequency $\omega_c$ where the transition changes between first-order to DQC. To find $\omega_c$, we assume $\omega_0 \gg 1$ such that most of the weight of the propagator will still be concentrated around $\tau=\tau'$. We can then Taylor expand
\begin{equation}
    \cos\{2[\phi(\tau)+\phi(\tau')]\} = \cos[4\phi(\tau)] + \text{irrel.}
\end{equation}
up to terms irrelevant in the parameter range $1/4<K<1/2$. At $K=1/4$, the next term in the Taylor expansion $\partial_\tau\phi\sin(4\phi)$ would become relevant. Due the linearization of the fermionic dispersion in bosonization and, as a consequence, the infinitely filled Fermi sea, fluctuations of $\phi$ can in principle be infinite. Therefore, in order to Taylor expand the difference, we have to normal-order first. Making use of the identity \cite{Knops1980,Nozieres1987}
\begin{equation}
    \cos(\phi) = :\cos(\phi):e^{-\frac{1}{2}\expval{\phi^2}_0}\,,
\end{equation}
where $:\cos(\phi):$ denotes the normal-ordered cosine and $\expval{.}_0$ denotes the vacuum expectation value with respect to the free part of \eqref{eq:SG}, the expansion becomes
\begin{equation}
    \cos\{2[\phi(\tau)-\phi(\tau')]\} = \left\{1-2(\tau-\tau')^2[\partial_\tau \phi(\tau)]^2\right\}e^{-2\expval{\left[\phi(\tau)-\phi(\tau')\right]^2}_0}\,.
\end{equation}
Such correlation functions have been determined in full generality \cite{Giamarchi2003}. In our case, we find
\begin{equation}
    \expval{\left[\phi(\tau)-\phi(\tau')\right]^2}_0 = K\log\abs{\tau-\tau'}
\end{equation}
such that the expansion reads
\begin{equation}
    \cos\{2[\phi(\tau)-\phi(\tau')]\} = \left\{1-2(\tau-\tau')^2[\partial_\tau \phi(\tau)]^2\right\}\abs{\tau-\tau'}^{-2K}\,.
\end{equation}
Inserting this back into our original action \eqref{eq:SGafterIntOut}, performing the $\tau'$ integration and rewriting the action back in Sine-Gordon form gives
\begin{align}
    \begin{split}
    S(\omega_0) =\int \dd{x}\dd{\tau} \bigg\{&\frac{1}{2\pi K_{\text{new}}}\left[\gamma (\partial_x\phi)^2+\gamma^{-1}(\partial_\tau\phi)^2\right]\\ &+ \mu_{\text{new}}\cos(4\phi)\bigg\}
    \end{split}
\end{align}
with new Luttinger parameter $K_{\text{new}} = \gamma K$ and  
\begin{align}
    \gamma &=\left(\frac{2\pi Kg^2u^{-2K}\Gamma(3-2K)}{\rho\omega_0^{4-2K}}+1\right)^{-\frac{1}{2}}\,,
\end{align}
where $\Gamma(x)$ is the Euler gamma function. Just like in the original Sine-Gordon model, this theory hosts a line of DQCPs for $\mu_{\text{new}} = 0$ and $\frac{1}{8}<K_{\text{new}}<\frac{1}{2}$. The Taylor expansion itself holds for 
$\frac{1}{4}<K<\frac{1}{2}$, as explained above. Both conditions can be fulfilled simultaneously, at least for some values of $K$, if $1/4<\gamma<2$. The upper bound is trivially obeyed. For the lower bound, we find the condition
\begin{equation}
    \rho \omega_0^{4-2K} > \frac{2\pi K \,\Gamma(3-2K)}{15}g^2
\end{equation}
which has to hold for DQC to persist even in the presence of phonons. Rewritten as a simple scaling relation, we find the critical scaling
\begin{equation}
     \omega_c^{4-2K} \sim g_c^2/ \rho\,.
\end{equation}
One can confirm this result agrees with Sec.~\ref{sec:DQCP1d} when we take the adiabatic limit $\omega\to 0$ for a fixed $\mathcal{K}$, giving $g_c^2/\mathcal{K} \sim \omega^{2-2K}_0 \to0$.

\subsection{Phonons In Two Dimensions}
Just like in 1d, we integrate out the phonons, yielding a retarded monopole-monopole interaction on top of $\mathcal{L}_c$ in \eqref{eq:NCCP1}, of the form
\begin{equation}
    S_{\Phi\Phi} = -2g^2\int \dd[d+1]{x} \dd[d+1]{y}\, \sum_{i=1}^2\Phi_i(x) G(x-y) \Phi_i(y)
    \label{eq:IntPhonons}
\end{equation}
where we have written $x=(\tau_x,\bm{x})$ and $y=(\tau_y,\bm{y})$. The phonon propagator is given by
\begin{equation}
    G(x-y) = \frac{1}{2\rho\omega_0}e^{-\omega_0\abs{\tau_x-\tau_y}}\delta(\bx - \by)\,.
    \label{eq:GreensFuncPhonons}
\end{equation}
Again, we consider first the adiabatic ($\omega_0\to 0$) and antiadiabatic ($\omega_0 \to \infty$) limits separately. The adiabatic case has been studied above. In the antiadiabatic case, we can make use of the delta-function identity \eqref{eq:AntiadiabaticDelta}. 
Using the mapping between monopoles and the VBS order parameter in the NLSM, this coupling generates an additional anisotropy
\begin{equation}
    \mathcal{L}^{\mathrm{NLSM}}_2[\tilde{\bm{n}}] + s(\bn^2-\bm{v}^2) - \frac{2g^2}{\rho\omega_0^2}(v_1^2+v_2^2)
\end{equation}
At first, it might seem like this destroys the emergent symmetry at the critical point $s=0$. But observe that we can rewrite
\begin{equation}
    s(\bn^2-\bm{v}^2) + \frac{2g^2}{\rho\omega_0^2}(v_1^2+v_2^2) = \left(s+\frac{g^2}{\rho\omega_0^2}\right)(\bn^2-\bm{v}^2) - \frac{g^2}{\rho\omega_0^2}(\bn^2+\bm{v}^2)\,.
\end{equation}
From $(\bn^2 + \bm{v}^2) = 1$, we find that in the antiadiabatic limit, the transition is simply shifted by
\begin{equation}
    s_c = 0 \overset{\omega_0 \gg 0}{\longrightarrow} s_c =-\frac{g^2}{\rho\omega_0^2}\,,
    \label{eq:ScShift}
\end{equation}
as in the 1d case.
Now, let us turn to the case in between where the transition changes from first order to DQC. Observe first that we can rescale the theory by $x_i = \frac{x'_i}{\omega_0}$
to move all dependence on $\omega_0$ into the coupling. The interaction now reads
\begin{equation}
    S_{\phi\phi} = \frac{g^2}{\rho_0\omega^{5-2\Delta_\Phi}}\int 
    \dd[2]{\bx}\dd{\tau_x'}\dd{\tau_y'} \sum_{i=1}^2\Phi_i(\tau_x')e^{-\abs{\tau_x'-\tau_y'}}\Phi_i(\tau_y')\,,
    \label{eq:NewInt}
\end{equation}
where from now on we suppress the space-dependence of the monopole.
Now we can perform a formal expansion of the monopole 
\begin{align}
    \begin{split}
    \Phi_i(\tau_y') = &\Phi_i(\tau_x') + (\tau_y'-\tau_x')\partial_{\tau_x'} \Phi_i(\tau_x')\\ &+ \frac{1}{2}(\tau_y'-\tau_x')^2\partial_{\tau_x'}^2 \Phi_i(\tau_x') + \order{\tau_x'^3}\,.
    \label{eq:TaylorExpMonopole}
    \end{split}
\end{align}
and perform the $\tau_y'$ integration to find
\begin{equation}
    \sim \frac{2g^2}{\rho_0\omega^{5-2\Delta_\Phi}}\int d^2\bx'\int d\tau'_x \sum_{i=1}^2 \left[\Phi_i^2 - (\partial_{\tau'_x} \Phi_i)^2\right]\,.
    \label{eq:NewTermFromPhonons}
\end{equation}
The first term simply generates an anisotropy, just like in the $\omega_0\to\infty$ limit. At first glance, the form of the anisotropy is different than in \eqref{eq:ScShift}. However, we have to keep in mind that the coordinates in \eqref{eq:NewTermFromPhonons} have been rescaled. In order to add this to the anisotropy in the NLSM, we must revert back to the original coordinates resulting in the correct shift \eqref{eq:ScShift}.
The second term simply modifies the kinetic term in the NLSM description. We therefore believe that DQCP will break down when this perturbative treatment breaks down and higher orders in the Taylor expansion \eqref{eq:TaylorExpMonopole} have to be included; this is $2g^2/(\rho\omega_0^{5-2\Delta_\Phi}) \sim 1$.
This suggests a critical scaling $g_c^2 \sim \rho\omega_c^{5-2\Delta_\Phi}$. This scaling argument can be performed in general dimension, revealing a critical 
\begin{align}
    g^2_c \sim \rho\omega_c^{3+d-2\Delta_{\Phi}}
    \quad \longleftrightarrow \quad
    g^2_c \sim \mathcal{K} \omega_c^{1+d-2\Delta_{\Phi}}
    \,.
    \label{eq:CritScalingGen}
\end{align}
This is consistent with the results obtained in 1d (where $\sin(2\phi)$ plays the role of $\Phi$), and is also compatible with an alternative CFT-based stability analysis of the $d=2$ DSL \cite{seifert2023spinpeierls}.

\begin{figure}
    \centering
    \includegraphics[width=0.9\columnwidth]{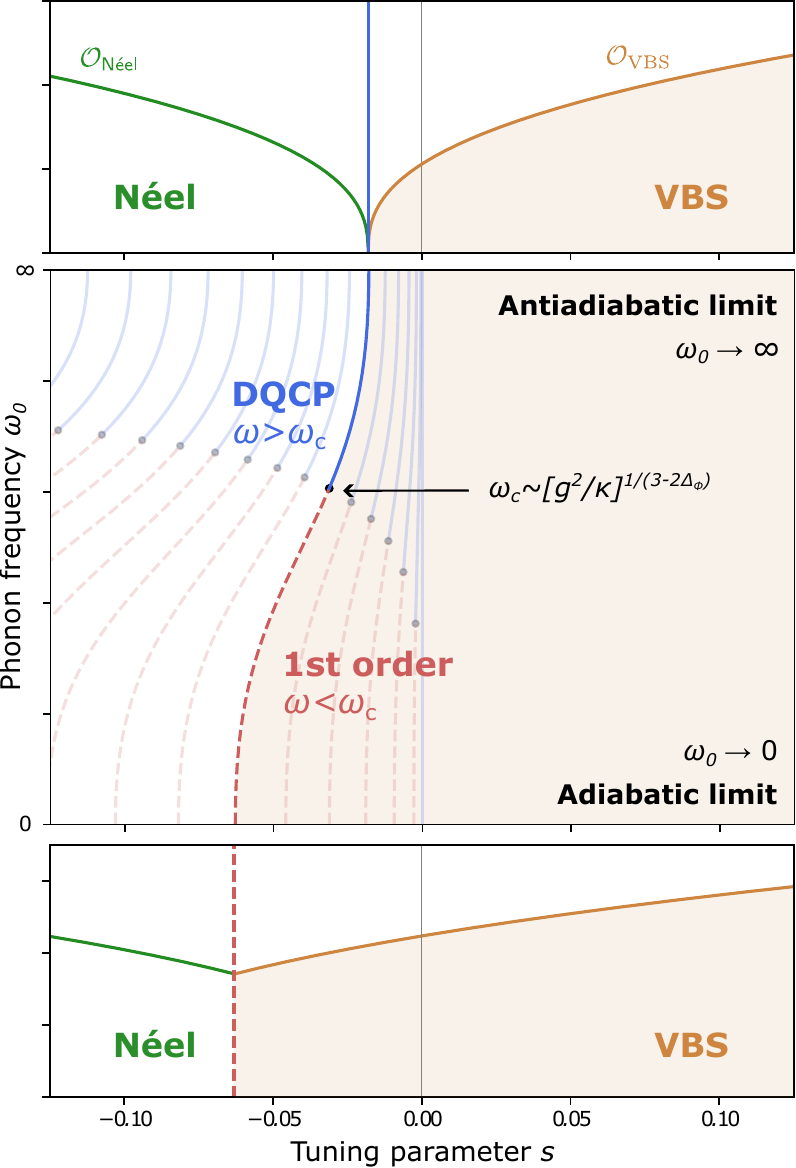}
    \caption{Nature of the Néel--VBS phase transition on the square lattice as a function of phonon frequency $\omega_0$, with $\mathcal{K}=\rho\omega_0^2$ constant. For $\omega_0\to\infty$ (the antiadiabatic limit, top panel), the critical $s_c$ is shifted but the transition remains second order.
    This DQCP persists down to $\omega_c$, below which it becomes first order.
    For $\omega_0\to 0$ (the adiabatic limit, bottom panel), the transition occurs at $s_{\mathrm{static}}$, given by Eq.~\eqref{eq:sstatic}.
    Shown here is $g = 0.3$, $\mathcal{K}=1$, with a range of $g=0.0$--$1.0$ shown in the background.
    }
    \label{fig:combinedfig}
\end{figure}

We summarize our results for the stability of the DQCP in Figure~\ref{fig:combinedfig} for a range of frequencies $\omega$. We simultaneously take the adiabatic $\omega\to0$ and antiadiabatic $\omega\to\infty$ limits by keeping $\mathcal{K} = \rho \omega_0^2$ constant. This choice to take $\rho\to0$ as $\omega_0\to\infty$ means that the shift of the transition \eqref{eq:ScShift} in the antiadiabatic limit remains finite.
We highlight that the $\omega_0$-dependent shift in the new transition point implies that a DQCP can be tuned \emph{by varying phonon frequency}, as has been seen in recent numerical work \cite{goetz2024phases}. In these models, increasing spin-phonon coupling or tuning parameters to push the transition to lower frequencies should make the transition strongly first order.

\section{Application to other systems}
While the existence of a true DQCP on the original square lattice model is still under debate \cite{Nahum2015,Poland2019,Ma2020,Nahum2020,DEmidio2023,takahashi2024so5}, our results may be expected to hold also for a scenario of a weakly-first order regime and approximate criticality.
Further, general underlying principles and the field-theoretic formulation of our results allows us to extend to several other closely related strongly-coupled gapless, \emph{deconfined} gauge theories where conformal symmetry may be present. 

In a previous work by some of us Ref.~\cite{seifert2023spinpeierls}, we have established that $\Uone$ DSL can exhibit spin-Peierls type instabilities upon coupling to lattice distortion modes, by exploiting a description of the low-energy field theory for the DSL (emergent quantum electrodynamics in 2+1 dims.) in terms of a strongly coupled conformal field theory with an emergent $\SU(4)$ symmetry. Note that it has been proposed that the low-energy theory of spinons strongly interacting with an emergent $\Uone$ gauge field possesses an equivalent description of a NLSM with a level-1 WZW term \cite{PhysRevB.97.195115} (see Appendix~\ref{app:DSL}).

The $U(1)$ DSL as a phase is intrinsically stable if there are no relevant operators which transform as singlets under the embedding of microscopic (UV) spin rotation and lattice symmetries $G_\mathrm{UV} \subset G_\mathrm{IR}$ \cite{song19,song20}.
Crucially, on the triangular and Kagome lattice, this embedding shows that there are six monopole operators of the emergent gauge field, three of which transform as VBS order parameters.
The latter have symmetry-allowed coupling to lattice distortion modes, and their strong relevance ($\Delta_\Phi < 3/2$) implies that the system exhibits a spin-Peierls instability, in analogy to the instability of the 1-dim. Luttinger liquid phase of spin chains \cite{seifert2023spinpeierls}.

The intrinsic stability of the $U(1)$ DSL as a phase on triangular and Kagome lattices relies on the assumption that (absent of perturbations) there are no relevant $G_\mathrm{UV}$-singlet operators, as indicated by recent analytical and numerical results \cite{chester16,alba22,song19,song20}.
In \cite{PhysRevB.97.195115} it was shown that, if a symmetry-allowed four-fermion interaction term $u(\overline\psi M \psi)^2$ were relevant,
it would produce a DQCP on the triangular lattice which tunes between 120-degree Néel and VBS order \cite{PhysRevB.97.195115}.
Following the general arguments in our work, one can write down a coupling of the VBS monopoles to the lattice as $-\bm{h}\cdot\bm{n}$
where $\bm{h} = g(u_1,u_2,u_3,0,0,0)$ has components given by three lattice distortions which are related by $C_3$ rotation symmetry.
Given the strong relevance of monopoles, we hence expect the destruction of deconfined criticality towards a strongly-first order transition. In the case of dynamical phonons, the second order transition is restored above frequencies $\omega_c^2 \sim [g^2/\mathcal{K}]^{3-2\Delta_\Phi}$.

More recently, it has been argued that QED$_3$ provides an effective theory for ``unnecessary'' quantum critical points and the surrounding critical regime \cite{zhang24}, by starting with the observation that the $U(1)$ DSL on the square lattice admits a G$_\mathrm{UV}$-singlet relevant monopole deformation $\Phi_2$. Tuning its coupling $\lambda$ therefore can be understood as a DQCP-like control parameter.
Crucially, in the scenario at hand, the nature of the resulting phase is argued to be independent of $\lambda$, and therefore $\lambda = \lambda_c$ may be a critical point \emph{within the same phase}.
In addition to this singlet monopole, the $\Uone$ DSL on the square lattice admits three monopole operators transforming as Néel order parameters as well as two monopole operators that transform as VBS order parameters.
Depending on the sign of an additional dangerously irrelevant coupling (similar to the DQCP anisotropy parameter), the unneccessary critical point can lie within the Néel or VBS order phases.

Adding spin-lattice couplings, the VBS monopole can be expected to couple to the distortion modes $u_{1,2}$, which are precisely those also discussed in Eq.~\eqref{eq:DistFieldGeneral}. By the arguments in the work at hand (and in Ref.~\cite{seifert2023spinpeierls}), the strong relevance of monopoles \cite{alba22} leads us to expect that the system spontaneously generates a lattice distortion in the adiabatic limit (i.e. classical lattice distortion modes).
The impact on the unnecessary critical point depends on the surrounding phases:
(i) In the case of the VBS--VBS critical point, it is plausible that the transition will be generically removed; instead of the VBS gap going to zero with a power law $\beta=\nu_\Phi\Delta_\Phi$, it will have a finite minimum on the order of $\Delta_0\sim (g^2/\mathcal{K})^{1/(3-2\Delta_\Phi)}$.
(ii) For the Néel--Néel critical point, we hypothesize that the second order transition would split into two first order Néel--VBS transitions. There would be an island of VBS order within the Néel phase, induced by the interaction of the lattice with strongly relevant monopoles. This intermediate VBS order would exist for $|\lambda| < [g^2/\mathcal{K}]^{(3-\Delta_\Phi)/(3-2\Delta_\Phi)}$.
In the antiadiabatic limit $\rho\to\infty$, we deduce that the unnecessary DQCP is preserved, which holds at finite-$\omega_0$ above to a critical frequency $\omega_c^2\sim [g^2/\mathcal{K}]^{3-2\Delta_\Phi}$ (see Appendix~\ref{app:unnecessary}).

The aforementioned 2d models fall into the recently proposed class of Stiefel liquids \cite{Zuo2021}. This work described the DQCP and $\Uone$ DSL as the $N=5$ and $N=6$ realizations of a series of critical models with an emergent $\SO(N)$ symmetry and critical behavior controlled by a WZW NLSM.
These phases are disordered critical states which may arise between competing VBS and (increasingly complex forms of) antiferromagnetic orders.
These therefore generalize the intertwinement originally proposed in 1d spin chains and the DQCP \cite{Senthil2006}.
In the case of the proposed non-Lagrangian $N=7$ Stiefel liquid, an analysis which relies on anomaly matching proposes that the four-component VBS order parameters proximate to the disordered phase transform on the triangular lattice with momenta $\bm{K}+\bm{M}.$ We identify the relevant perturbation $-\bm{h}\cdot \tilde{\bm{n}}$ to appropriate commensurate lattice distortion modes $\bm{h} = g(0,0,0,u_1,u_2,u_3,u_4)$. Given that the $\SO(7)$-vector perturbation $\tilde{\bm{n}}$ is potentially highly relevant \cite{he2022}, this could indicate a spin-Peierls instability for this phase too.
Our results may be understood as a natural obstacle to the intertwinement of VBS and antiferromagnetic orders in such a way. Any Stiefel liquids which exist between such competing orders could thus be unstable (under the assumption of the adiabatic limit/classical distortion modes) to VBS ordering.

\section{Conclusion}
In this article, we have outlined conditions regarding the stability of the DQCP under coupling to lattice degrees of freedom.
The instability occurs for infinitesimal couplings when considering static distortions, while finite phonon frequencies protect the system up to some nonzero critical spin-lattice coupling.
Hence, gapless systems that admit strongly relevant instantons which precipitate VBS order may be susceptible to a spin-Peierls distortion upon coupling to the lattice, in particular also in higher-dimensional systems (in addition to the familiar one-dimensional examples).
For the DQCP to remain stable, the spin-lattice coupling must be smaller than some critical value that depends on the phonon frequency. The scaling relations obtained here are consistent with the ones obtained for the $U(1)$ DSL in \cite{seifert2023spinpeierls}.
Though, of course, the exact microscopic instability is lattice- and even model-dependent.

From a field theoretic point of view, this possible instability relies on the existence of a strongly relevant but lattice symmetry-breaking operator which can couple to an appropriate lattice distortion mode. 
Both for the DQCP and the DSL, these operators are the monopoles of the corresponding gauge theory, but our formalism indicates that this mechanism holds more generally if: (i) the system (in the absence of a distortion) is gapless and (ii) the energy gain due to the gap opening stemming from the distortion coupling to the lattice-symmetry breaking operator $\mathcal{O}$ scales with ${\chi = \frac{d+1}{d+1-\Delta_{\mathcal{O}}}<2}\,$, where $\Delta_{\mathcal{O}}$ is the scaling dimension of $\mathcal{O}$ and $d$ is the spatial dimension [see Eq.~\ref{eq:energycompetition} in the Introduction].
Such a lattice symmetry breaking operator often exists in quantum magnets as a (fluctuating) VBS order parameter.
Our field-theoretic arguments primarily rely on UV symmetry quantum numbers as well as minimal assumptions concerning the structure of the IR critical theory of the states in question.
It is interesting to verify our predictions in numerical simulations of microscopic models. Recent numerical DMRG and Gutzwiller-projected variational Monte Carlo simulations indicate a strong propensity of the gapless $\Uone$ DSL to undergo a spin-Peierls instability on the triangular \cite{seifert2023spinpeierls} and Kagome lattices \cite{Ferrari2024}, respectively.

Quantum Monte Carlo simulations of lattice models for DQCP in 2d are notoriously hard due to the existence of a sign problem.
Remarkably, it has recently been pointed out that a sign-problem free formulation of the Su--Schrieffer--Heeger-Hubbard model realizes an effective unconstrained $\Ztwo$ gauge theory, with quantum Monte Carlo simulations providing evidence for a direct continuous Néel--VBS transition \cite{kara18,goetz2024phases}.
Upon tuning a control parameter in the lattice model which effectively maps onto a ratio of phonon frequency and spin-lattice coupling, the continuous transition can be tuned to become strongly first order.
It will be highly desirable to quantitatively extract the shift of the location of the DQCP as well as the regime of stable criticality as a function of spin-lattice coupling and phonon frequency. Given limited system sizes and the required double scaling, this is expected to be a more challenging task in 2d. In 1d, however, many numerical studies of quantum spin chains coupled to phonons exist, see, e.g., \cite{PhysRevResearch.2.023013,weber2023,Uhrig1998,Ferrari2020} and references therein; so we expect it to be possible to extract the aforementioned features.
The extracted critical exponents in higher dimension would provide interesting insight into the nature of the DQCP and the field theory's operator content.

Turning towards experiments, a recent study observed ``proximate-DQCP'' behavior between plaquette-solid and AFM phases in SrCu$_2$(BO$_3$)$_2$ under tuning of the magnetic field \cite{Cui2023}.
The regime of quantum-critical scaling observed here was not reproduced in another work which instead tunes pressure to induce the transition \cite{Guo2023}. Varing pressure is a standard technique to tune magnetic couplings, but inherently also changes the phonon energy scale. Due to the signigicant interplay of spin and lattice degrees of freedom in candidate DQCP materials, the strong first-order transition here does not necessarily exclude the potential of observing an underlying DQCP by tuning lattice-independent parameters. 
We emphasize, as has been previously suggested \cite{mila23}, that clarifying the interplay of spin and lattice degrees of freedom in such system may be crucial for a more complete understanding.
Our work at hand may thus constitute a step \cite{Lee2019} towards understanding the conditions for deconfined quantum criticality in realistic experiments.

\acknowledgements

We gratefully acknowledge discussions with F. Assaad, A. Chubukov and B. Douçot.
U.F.P.S. acknowledges support from the Deutsche Forschungsgemeinschaft (DFG, German Research Foundation) through a Walter Benjamin fellowship, Project ID~449890867. This research was supported in part by the National Science Foundation under Grant No. NSF PHY-1748958.
J.K. acknowledges support from the Imperial-Technical University of Munich flagship partnership and financial support by the Deutsche Forschungsgemeinschaft (DFG, German Research Foundation) via TRR 360 (Project-ID No. 492547816). The research is part of the Munich Quantum Valley, which is supported by the Bavarian state government with funds from the Hightech Agenda Bayern Plus. 

\appendix

\section{NLSM for the DSL}\label{app:DSL}

It has been proposed that the Dirac spin liquid has a complementary description in terms of a $\SO(6)$ NLSM with a WZW term \cite{Senthil2006,PhysRevB.97.195115,Zuo2021}.
Such theory is constructed by considering a low-energy theory for fluctuations of a chiral flavor-symmetry breaking order parameter $\mathcal{P}_{ij}$ which couples as a mass for the fermions, $m \,\mathcal{P}_{ij} \, \overline{\psi}_i \psi_j$. Skyrmions of the field $\mathcal{P}$ are charged under the dynamical gauge field $a$ and thus correspond to fermions in the original theory.
The effective theory of $\mathcal{P}_{ij}$ is a NLSM with a level-1 WZW term \cite{PhysRevB.97.195115}.
Writing in the basis of adjoint $\SO(6)$ generators $\sigma_i\tau_j$ (where $i=0$ is the identity matrix and $\sigma,\tau$ are Pauli matrices which act on the spin/valley degrees of freedom), introduce the $\SO(6)$ order parameter field $\bm{n}$ through $\mathcal{P}_{ij} = - 2\iu \bm{n}^T \sigma_i\tau_j \bm{n}$, leading to the effective theory
\begin{equation}
    S_2^{\mathrm{DSL}}[\bm{n}] =  S_2^{\mathrm{kin}}[\bm{n}]
    + \Gamma_2^{\SO(6)}[\bm{n}]\,.
    \label{eq:sigmamodel_o6}
\end{equation}
A careful treatment uncovers that the vector $\bm{n}$ couples linearly to $A_{\mathrm{top}}$ and thus correspond to the monopoles in the original theory \cite{Zuo2021}.

\section{Unnecessary critical points}\label{app:unnecessary}
Here, for completeness, we briefly review recent arguments of Ref.~\cite{zhang24}.
The monopoles $\Phi_{4,5,6}$ of the $U(1)$ on the square lattice form a triplet under spin-rotation symmetry and therefore transform as Néel order parameters $N_{x,y,z}$, just as on the triangular lattice.
Of the remaining three monopoles, $\Phi_2$ transforms trivially under the UV symmetry group, and $\Phi_{1,3} \sim v_{1,2}$ transform as columnar VBS order parameters.
Upon the proliferation of the trivial monopole for $|\lambda| > 0$, the global symmetry is reduced $\SO(6)\times \Uone_{\mathrm{topo}} \to \SO(5)$ \cite{song19,song20}.
The transformation of these uncondensed monopoles corresponds to the fermion bilinears studied earlier in \cite{hermele2005}. 
These order parameters can be written as a 5-component NLSM $\tilde{\bm{n}}$ and the effective theory of the system is 
\begin{equation}
    S_2^{\mathrm{kin}}[\tilde{\bm{n}}] + \Gamma^{\SO(5)}_2[\tilde{\bm{n}}] - \int\!\dd[3]{x} \kappa (\bm{n}^2 - \bm{v}^2),
\end{equation}
where $\kappa$ represents an anisotropy which further breaks $\SO(5)\to\SO(3)\times U(1)$. Although it is irrelevant at the original $\SO(6)$-symmetric QED$_3$ point $\lambda=0$ (since it corresponds to double-strength monopoles), it is expected to be relevant about the new $\SO(5)$-symmetric NLSM FP (here it is really the original anisotropy of the DQCP). 
This is therefore a dangerously irrelevant coupling at $\lambda=0$, the sign of which determines the nature of the ordered phase on both sides of the transition.

If it is energetically favorable for the lattice to distort, then the $\SO(2)$ VBS group, which is promoted to $\SO(6)\times \Uone$ at $\lambda=0$, is broken down to $\Z_4$.
In the antiadiabatic limit, the (marginally irrelevant) coupling is shifted as in Eq.~\eqref{eq:ScShift}, by $\kappa \to \kappa- g^2 / \rho\omega_0^2$. Hence as $\rho\to\infty$ the unnecessary DQCP is preserved but the VBS--VBS transition is favored.

\bibliographystyle{apsrev4-2}
\bibliography{main}

\end{document}